\documentclass[12pt]{article}%
\usepackage{amsmath,amsfonts,amssymb,amsthm,amstext,amscd,eucal}
\usepackage{latexsym}
\usepackage{sparticles}
\usepackage{feynmf}
\usepackage[dvips]{graphicx}
\usepackage{amsmath}
\usepackage{amsfonts}
\usepackage{amssymb}%
\makeatletter \@addtoreset{equation}{section}

\makeatletter\renewcommand\section{\@startsection {section}{1}{\z@}                                   {-3.5ex \@plus -1ex \@minus -.2ex}                                   {2.3ex \@plus.2ex}                                   {\normalfont\large\bfseries}}
\renewcommand\subsection{\@startsection{subsection}{2}{\z@}                                     {-3.25ex\@plus -1ex \@minus -.2ex}                                     {1.5ex \@plus .2ex}                                     {\normalfont\bfseries}}
\newcommand{\be}{\begin{equation}}
\newcommand{\ee}{\end{equation}}
\newcommand{\bea}{\begin{eqnarray}}
\newcommand{\eea}{\end{eqnarray}}

\newcommand{\bse}{\begin{subequations}}
\newcommand{\ese}{\end{subequations}}
\newcommand{\bi}{\begin{itemize}}
\newcommand{\ei}{\end{itemize}}

\newcommand{\mpl}{M_{\rm pl}}
\newcommand{\Ld}{\Lambda_{\rm dressed}}

\begin{document}
\begin{titlepage}
\begin{flushright}\vspace{-3cm}
{\small
UUITP-26/11,\\
IPM/P-2011/052 \\
\today }\end{flushright}
\vspace{-.5cm}
\begin{center}
\centerline{{\Large{\bf{$1/N$ Resolution to Inflationary $\eta$-Problem }}}}
\vspace{2mm}
{\large{{\bf A.~Ashoorioon\footnote{Email: amjad.ashoorioon@fysast.uu.se}$^{,a}$,  U. Danielsson\footnote{Email: ulf.danielsson@physics.uu.se}$^{,a}$,
M.M. Sheikh-Jabbari\footnote{Email:
jabbari@theory.ipm.ac.ir}$^{,b}$}}}
\\
\vspace{5mm}
\bigskip\medskip
\begin{center}
$^a$ {\it Institutionen f\"{o}r fysik och astronomi
Uppsala Universitet,\\ Box 803, SE-751 08 Uppsala, Sweden}\\
\smallskip
$^b$ {\it School of Physics, Institute for Research in Fundamental
Sciences (IPM),\\ P.O.Box 19395-5531, Tehran, Iran}\\
\end{center}%
\end{center}
\setcounter{footnote}{0}
\date{\today}
\begin{abstract}
We observe that the dominant one loop contribution to the graviton propagator in the theory of $N$ ($N\gg 1$) light scalar fields $\phi_a$ (with masses smaller than $\mpl/\sqrt{N}$) minimally coupled to Einstein gravity is proportional to $N$ while that of  graviton-scalar-scalar interaction vertex is $N$ independent. We use this to argue that  the coefficient of the $R\phi_a^2$ term appearing at one loop level is $1/N$ suppressed. This observation provides a resolution to the $\eta$-problem, that the slow-roll parameter $\eta$ receives order one quantum loop corrections for inflationary models built within the framework of scalar fields minimally coupled to Einstein gravity,  for  models involving large number of fields. As particular examples, we employ this to argue in favor of the absence of $\eta$-problem in M-flation and N-flation scenarios.
\end{abstract}
\end{titlepage}\renewcommand{\baselinestretch}{1.1}


\section{Introduction}

Recent CMB observations \cite{arXiv:1001.4538} indicate that the early
Universe has passed through an inflationary period with Hubble parameter
$H\lesssim10^{-5}M_{\mathrm{pl}}$. The standard theoretical setup for
inflationary models generically involve some scalar inflaton fields slowly
rolling down their potential. The slow-roll that is needed to ensure a
resolution to the problems of standard big bang cosmology, and consistency
with the CMB results, demands $\epsilon=-\frac{\dot{H}}{H^{2}}$ and
$\eta=\frac{\ddot{H}}{H\dot{H}}$ to be smaller or of order $10^{-2}$
\cite{arXiv:1001.4538}. In the context of simple single scalar models with
potential $V(\phi)$, $\epsilon$ and $\eta$ are a measure of flatness of the
potential and, specifically, $\eta\sim m^{2}/H^{2}$, where $m$ is the
effective mass of the inflaton field. Therefore, the theoretical framework
invoked for inflationary model building should also provide mechanisms to
protect the potential and its flatness against quantum and/or quantum gravity
corrections. In physics models we generically associate smallness and
protection of a quantity like $\eta$ with an approximate symmetry, such that
when the symmetry is exact this parameter (here the effective mass of
inflaton) is zero. Supersymmetry, for example, can be such a symmetry. It
turns out that in the presence of gravity, as in our case where we are dealing
with inflation models, symmetries protecting $\eta$ are all broken, inducing an
inflaton mass term of order the Hubble scale $H$ of the background. We hence
end up with an order one $\eta$, i.e., the $\eta$-problem.

From the above discussion it is seen that the $\eta$-problem may appear in two
ways: In a top-down approach where we invoke a theory of quantum gravity like
string theory for inflationary model building. Or in a bottom-up approach
where we take the usual field theory setup of Einstein gravity plus scalar
inflaton fields, assuming that this framework is valid up to Planck scale
$M_{\mathrm{pl}}$. In the top-down approach the $\eta$-problem appears as a
classical (not loop) effect, usually due to the interaction of the
inflationary sector with the \textquotedblleft moduli stabilization
sector\textquotedblright, see, e.g., \cite{Stringy-eta}. In these models it
turns out to be easy to have small $\epsilon$ with controlled back-reactions
on it, but $\eta$ receives order one corrections. Intuitively, in these
top-down setups, the $\eta$-problem can be understood as follows. With a
vacuum energy of order $V$ all scalars, including the inflaton, will be
endowed with soft masses of order $V/M_{\mathrm{pl}}^{2}=3H^{2}$ since the
supersymmetry breaking scale is not below $H$.
In this work, however, we will focus on the $\eta$-problem in the bottom-up approach.

In the bottom-up approach the $\eta$-problem arises from \textit{quantum} loop
corrections to the tree-level graviton-scalar-scalar vertex.
Despite being non-renormalizable, one can still apply the (Wilsonian) effective field
theory techniques to the Einstein gravity theory and consider loop
corrections, e.g. see \cite{Gravity-loops,Duff:1974hb,'tHooft:1974bx}
. In the presence of a scalar field $\phi$ minimally coupled to Einstein
gravity, as in generic inflationary models, these corrections at one loop
level generically involve a $R\phi^{2}$ term, a term whose presence was noted
long ago \cite{Sakharov-Adler}. As we will review below, such a term is
quadratically divergent and in the one loop effective action appears as
\begin{equation}
\zeta\frac{\Lambda^{2}}{M_{\mathrm{pl}}^{2}}R\phi^{2}, \label{conf-mass-zeta}%
\end{equation}
where $\Lambda$ is the UV cutoff of the theory and $\zeta$ is an order one
coefficient. Assuming a Planckian cutoff scale, $\Lambda\sim M_{\mathrm{pl}}$,
in an inflationary background where $R\sim H^{2}$, this term yields a
correction of order $H^{2}$ to the inflaton mass, causing the $\eta$-problem.


The $\eta$-problem, or the Hubble scale mass term for the effective inflaton
field, seems quite generic and one may put forward the idea of
\emph{kinematically} reducing the coefficient $\zeta\Lambda^{2}/M_{\mathrm{pl}%
}^{2}$.
In this Letter we explore this possibility. One obvious possibility is to
choose the cutoff $\Lambda$, the scale where quantum gravity effects become
important, to be one or two orders of magnitude smaller than $M_{\mathrm{pl}}$
\cite{hep-ph/9803315}. In this case the coupling constant of gravitons will be
reduced like the momentum UV cutoff and the $\eta$ problem persists.
Alternatively, one may explore the idea that $\zeta$ is a kinematical factor
that for some reason is not of order one, while the bare cutoff $\Lambda$ is
$M_{\mathrm{pl}}$. In fact, similar suppressions are already very well known
in the context of large $N$ gauge theories \cite{'tHooft:1973jz}: the
nonplanar part of a given Feynman diagram comes with powers of $1/N$
suppression compared to the planar part of the same diagram. As we will show
similar analysis can be repeated for the theories involving large number of
scalar fields minimally coupled to gravity. In particular, if we have $N$
number of \textquotedblleft light fields\textquotedblright, lighter than
$M_{\mathrm{pl}}/\sqrt{N}$, $\zeta$ turns out to have a $1/N$ suppression
factor. In a sense, as if, the diagram leading to \eqref{conf-mass-zeta} is a
nonplanar diagram. This observation is closely related to the
\textquotedblleft species dressed gravity cutoff scale\textquotedblright%
\ ideas discussed in \cite{Dvali:2007hz,Dvali:2008jb}, in light of which the
$\zeta\sim1/N$ result may be interpreted as dealing with a \textquotedblleft
dressed cutoff\textquotedblright\ $\Lambda/\sqrt{N}$ while $\Lambda\sim
M_{\mathrm{pl}}$.


Inflationary models with many scalar fields have recently got attention in
view of their success in providing a natural explanation for the smallness of
the inflaton self-couplings (the issue of steepness of the potential) and for
the super-Planckian excursion of the effective inflaton in the field space
\cite{Liddle:1998jc}.
This idea is not exotic to string theory motivated inflationary settings where
it is quite common to have an abundant number of fields/degrees of freedom
with masses below the dressed cutoff $M_{\mathrm{pl}}/\sqrt{N}$, see, e.g,
\cite{Nflation, Becker:2005sg,Mflation}. Even though in some of these setups,
like N-flation, the individual field excursion is greater than the dressed UV
cutoff, some, like Gauged M-flation \cite{gauged-M-flation} or multiple
M5-branes Inflation \cite{Becker:2005sg}, remain immune to
this\textquotedblleft beyond-the-cutoff\textquotedblright\ problem.

In this work we examine the above proposed $1/N$ resolution to the $\eta
$-problem. We assume that there is a hierarchy of scales between $H$, the
dressed gravity cutoff $\Lambda_{\mathrm{dressed}}\sim M_{\mathrm{pl}}%
/\sqrt{N}$ and $M_{\mathrm{pl}}$: $H\ll\Lambda_{\mathrm{dressed}}\ll
M_{\mathrm{pl}}$ which is easily achieved by e.g. $N\gtrsim10^{2}$. This
provides a window where one can safely use the standard techniques of quantum
field theory and effectively deal with a system that could be described by
Einstein-Hilbert gravity, the inflaton sector, and other heavy remnants of the
theory of quantum gravity whose masses $M_{a}$ falls below the new gravity
cutoff, i.e. $M_{a}\lesssim\Lambda_{\mathrm{dressed}}$.

The outline of this work is as follows. We consider a system of $N$ light
scalars minimally coupled to Einstein gravity and work out basic Feynman rules
of the theory and compute the quadratically-divergent part of the one loop
contributions to the graviton propagator and graviton-scalar-scalar vertices.
We show that one loop graviton two-point function has a linear $N$ parametric
dependence while the graviton-scalar-scalar vertex has no $N$ dependence.
Therefore, if we (re)normalize the graviton two-point function, the vertex
will have a factor of $1/\sqrt{N}$. This latter leads to $\zeta\sim1/N$
(\emph{cf.} \eqref{conf-mass-zeta}). We discuss how this can resolve the
$\eta$-problem in the context of many-field models like N-flation
\cite{Nflation} or M-flation \cite{Mflation,{gauged-M-flation}}.

\section{Loop analysis in multi-field inflationary model}

Consider the action of $N$ scalars minimally coupled to gravity
\begin{equation}
\mathcal{L}=-\frac{1}{2}M_{\mathrm{pl}}^{2}R-\frac{1}{2}\partial_{\mu}\phi
_{a}\partial^{\mu}\phi_{a}-\frac{1}{2}M_{a}^{2}\phi_{a}^{2}-V(\phi
_{a}),\label{action-scalar-field}%
\end{equation}
where $a=1,2,\cdots,N$ is the number of scalar fields, and summation over
repeated $a,b$ indices is assumed. One or some of these scalars play the role
of inflaton(s), and $V(\phi_{a})$ could be any potential that realizes
slow-roll inflation at the classical level, while the rest exhibit possible remnants of the underlying quantum
gravity theory. We assume the mass of these remnants to be below our dressed cutoff $\Ld$.
The action \eqref{action-scalar-field} once quantized will receive all
possible corrections compatible with the symmetries of the system, in
particular an $R\phi_{a}^{2}$ correction which appears at one loop level. As
discussed if this term comes with an order one coupling can cause the $\eta
$-problem. In what follows we show by carrying out explicit one loop
calculations involving gravitons, that this term is suppressed by factors of $1/N$, providing a
setting to resolve the $\eta$-problem in the context of multi-field models of inflation. In our analysis in this section and section 3 we ignore the loops involving scalar self-interactions. As we will discuss in the discussion section these diagrams do not change our main result.

\subsection{Tree level Feynman diagrams}

To perform the one loop analysis, as in any quantum field theory, we need to
work out basic tree level Feynman diagrams of propagators and interaction
vertices. To do so for the gravity sector, following \cite{Duff:1974hb}, we
introduce the tensor densities,
\begin{equation}
\bar{g}^{\mu\nu}=g^{1/2}g^{\mu\nu}\qquad\mathrm{and} \qquad\bar{g}_{\mu\nu
}=g^{-1/2}g_{\mu\nu},
\end{equation}
to bring the gravitational part of the action to the Goldberg's form
\cite{Goldberg:1958zz}
\begin{equation}
\frac{M_{\mathrm{pl}}^{2}}{16}\left(  2\bar{g}^{\rho\sigma}\bar{g}_{\lambda
\mu}\bar{g}_{\kappa\nu}-\bar{g}^{\rho\sigma}\bar{g}_{\mu\kappa}\bar
{g}_{\lambda\nu}- 4\delta^{\sigma}_{\kappa}\delta^{\rho}_{\lambda}\bar{g}%
_{\mu\nu}\right)  \bar{g}^{\mu\kappa}_{,\rho}\bar{g}^{\lambda\nu}_{,\sigma}.
\end{equation}
We will decompose the density metric to the flat part and the deviation from
the flat space part,
\begin{equation}
\label{metric-decomposition}\bar{g}_{\mu\nu}=\eta_{\mu\nu}+\hat{h}_{\mu\nu},
\end{equation}
where $\hat{h}_{\mu\nu}$ is defined as
\begin{equation}
\label{h}\hat{h}_{\mu\nu} \equiv M_{\mathrm{pl}}^{-1} h_{\mu\nu}.
\end{equation}
The inverse of $\bar{g}_{\mu\nu}$ is given by
\begin{equation}
\bar{g}^{\mu\nu}=\eta^{\mu\nu}-\hat{h}^{\mu\nu}+\mathcal{O}(\hat{h}^{2}),
\end{equation}
where on the R.H.S. the indices are raised and lowered by the flat Minkowski
space metric $\eta_{\mu\nu}$. Perturbing the action
\eqref{action-scalar-field} up to third order in $h_{\mu\nu}$, one obtains
\begin{equation}
\mathcal{L}=-\frac{1}{2} \partial_{\alpha}{h_{\mu\nu}}\partial^{\alpha}%
{{h}^{\mu\nu}}+\frac12\phi_{a}(\Box-M_{a}^{2}) \phi_{a} +\frac{1}%
{2M_{\mathrm{pl}}} h_{\mu\nu} T^{\mu\nu}+\frac{1}{M_{\mathrm{pl}}}%
\mathcal{O}({h}{(\partial{h}})^{2})
\end{equation}
where
\begin{equation}
\label{Tmunu}T^{\mu\nu}=\partial^{\mu}\phi_{a}\partial^{\nu}\phi_{a}-\frac
{1}{2}\eta^{\mu\nu}\partial^{\alpha}\phi_{a}\partial_{\alpha}\phi_{a}
+\eta^{\mu\nu}(\frac12 M_{a}^{2}\phi_{a}^{2}+ V(\phi_{a}))\,.
\end{equation}
Note that $T_{\mu\nu}$ is written to lowest order in $h_{\mu\nu}$ and so it is
independent of $h_{\mu\nu}$. 
\begin{figure}[t]
\begin{center}
\begin{fmffile}{grp} 	
  \fmfframe(0,0)(0,5){ 	
   \begin{fmfgraph*}(40,30)
\fmfleft{i1}\fmfright{o1}
\fmf{dbl_wiggly,label=$\vec p$,l.side=left}{i1,o1}
\fmflabel{$\mu\nu$}{i1}
\fmflabel{$\rho\sigma\quad=D_{\mu\nu\rho\sigma}=\frac{i\left(\eta_{\mu\rho}\eta_{\nu\sigma}
+\eta_{\mu\sigma}\eta_{\nu\rho}-\eta_{\mu\nu}\eta_{\rho\sigma}\right)}{p^2+i\epsilon}$}{o1}
 \end{fmfgraph*}
 }
\end{fmffile}\\
\begin{fmffile}{spr} 	
  \fmfframe(0,5)(0,5){ 	
   \begin{fmfgraph*}(40,30)
\fmfleft{i1}\fmfright{o1}
\fmf{dashes_arrow,label=$\vec p$,l.side=left}{i1,o1}
\fmflabel{$a$}{i1}
\fmflabel{$b~~\quad=\frac{i}{p^2+m^2_a}\delta^{ab}$}{o1}
 \end{fmfgraph*}
 }
\end{fmffile}\\
\begin{fmffile}{hpp0}
\fmfframe(0,10)(0,20){
 \begin{fmfgraph*}(80,80)
      \fmfleft{i1}
      \fmfrightn{o}{2}
       \fmf{dbl_wiggly,label=$\vec q$}{i1,w1}
      \fmflabel{$\mu\nu$}{i1}
      \fmflabel{$a$}{o1}
      \fmflabel{$b$}{o2}
      \fmfdot{w1}
      \fmf{dashes_arrow,label=$p^{\mu}$,label.side=left,label.dist=0.03w}{o1,w1}
      \fmf{dashes_arrow,label=$p^{\prime\nu}$,label.side=left,label.dist=1}{w1,o2}
\fmflabel{$~~~~~~~~~=V_{3\mu\nu}^{ab}(p,p^{\prime})\propto \frac{1}{\mpl}\delta^{ab}$}{w1}
    \end{fmfgraph*}
    }
 \end{fmffile}\\
\begin{fmffile}{hhpp} 	
\fmfframe(0,20)(0,20){
  \begin{fmfgraph*}(80,80)
      \fmfleft{i1,i2}
      \fmfright{o1,o2}
       \fmf{dbl_wiggly}{i1,w1}
      \fmflabel{$\mu\nu$}{i1}
      \fmf{dbl_wiggly}{i2,w1}
      \fmflabel{$\alpha\beta$}{i2}
      \fmflabel{$a$}{o1}
      \fmflabel{$b$}{o2}
      \fmf{dashes_arrow,label=$p$,l.side=left,l.d=0.05w}{o1,w1}
      \fmf{dashes_arrow,label=$p^{\prime}$,l.side=left,l.d=0.01w}{w1,o2}
\fmflabel{$~~~~~~~~~~=V_{4\alpha\beta\mu\nu}^{ab}(p,p^{\prime})\propto\frac{1}{\mpl^2}\delta^{ab}$}{w1}
     \end{fmfgraph*}
     }
\end{fmffile}\\
\begin{fmffile}{hhh0} 	
  \fmfframe(0,20)(0,0){
   \begin{fmfgraph*}(80,80)
      \fmfleft{i1}
      \fmfright{o1,o2}
       \fmf{dbl_wiggly,label=$\leftarrow p_1$}{w1,i1}
      \fmflabel{$\alpha\beta$}{i1}
      \fmfdot{w1}
      \fmf{dbl_wiggly, tension=0.5, label=$p_2 \searrow$,l.side=right,l.d=0.01}{w1,o1}
      \fmflabel{$\gamma\kappa$}{o1}
      \fmf{dbl_wiggly, tension=0.5, label=$p_3 \nearrow$,l.side=left,l.d=0.01}{w1,o2}
      \fmflabel{$\mu\nu$}{o2}
      \fmfv{label=$~~~~~=W_{3\alpha\beta\mu\nu\gamma\kappa}(p_2,,p_3)\propto\frac{1}{\mpl}$, label.angle=0, l.d=0.4w}{w1}
    \end{fmfgraph*}}\end{fmffile}
\caption{Some of tree level basic Feynman graphs of the theory relevant to our computations}
\label{hphiphi1}
\end{center}
\end{figure}
From this interaction term, and dropping the last term in
\eqref{Tmunu} which is inessential for our purposes, one can read the vertex
$\phi_{a}\phi_{a}{h}_{\mu\nu}$ to be $\frac{1}{M_{\mathrm{pl}}}\left(  p^{\mu}
p^{\prime\nu}-\frac{1}{2}\eta^{\mu\nu}(p\cdot p^{\prime})\right)  \delta_{ab}%
$, where $p^{\mu}$ and $p^{\prime\mu}$ are two external four-momenta on the
$\phi_{a}$ particles. To work out the basic Feynman graphs of the theory we
need to gauge-fix the diffeomorphism invariance. This may be done through
gauge fixing term
\begin{equation}
\mathcal{L}_{\mathrm{g.f.}}=\frac{M_{\mathrm{pl}}^{2}}{4}\bar{g}^{\mu\alpha
}_{,\mu}\bar{g}^{\nu\beta}_{,\nu}\eta_{\alpha\beta}.
\end{equation}
In Figure \ref{hphiphi1} we have plotted the basic Feynman graphs in this gauge.

\subsection{One loop analysis}

Having the tree level theory we now proceed to the one loop analysis and
revisit the one loop propagator calculations as well as graviton-scalar
vertex. Since we are interested in the quadratically divergent term
\eqref{conf-mass-zeta}, it is appropriate to use cutoff regularization;
dimensional regularization, which is very well suited in capturing the
logarithmic divergences and already used in
\cite{'tHooft:1974bx,Steinwachs:2011zs}, can not be employed here.

\subsubsection{One loop propagators}

\label{one-loop-prop}

\begin{figure}[tbh]
\begin{center}
\begin{fmffile}{hh1} 	
\fmfframe(10,10)(30,0){
\begin{fmfgraph*}(100,100)
\fmfleft{i1}
\fmflabel{$\mu\nu$}{i1}
\fmfright{o1}
\fmflabel{$\rho\sigma$}{o1}
\fmf{dbl_wiggly,label=$\vec p$}{i1,v1}
\fmf{dbl_wiggly,left,tension=0.4,label=$\vec k$,l.side=left}{v1,v2}
\fmf{dbl_wiggly,left,tension=0.4,label=$\vec p-\vec k$,l.side=left,l.d=0.1}{v2,v1}
\fmf{dbl_wiggly,label=$\vec p$}{v2,o1}
\fmfdot{v1,v2}
\end{fmfgraph*}
}
\end{fmffile}
\begin{fmffile}{hh2} 	
\fmfframe(30,0)(10,0){
\begin{fmfgraph*}(100,100)
\fmfleft{i1}
\fmflabel{$\mu\nu$}{i1}
\fmfright{o1}
\fmflabel{$\rho\sigma$}{o1}
\fmf{dbl_wiggly,label=$\vec p$}{i1,v1}
\fmf{dbl_wiggly,tension=1,label=$\vec k$,l.s=right}{v1,v1}
\fmf{dbl_wiggly,label=$\vec p$}{v1,o1}
\fmfdot{v1}
\end{fmfgraph*}
}
\end{fmffile}\newline%
\begin{fmffile}{hh3} 	
\fmfframe(10,10)(30,10){
\begin{fmfgraph*}(100,100)
\fmfleft{i1}
\fmflabel{$\mu\nu$}{i1}
\fmfright{o1}
\fmflabel{$\rho\sigma$}{o1}
\fmf{dbl_wiggly,label=$\vec p$}{i1,v1}
\fmf{dashes_arrow,left,tension=0.4,label=$k$,l.side=left}{v1,v2}
\fmf{dashes_arrow,left,tension=0.4,label=$k-p$,l.side=left,l.d=0.1}{v2,v1}
\fmf{dbl_wiggly,label=$\vec p$}{v2,o1}
\fmfdot{v1,v2}
\end{fmfgraph*}
}
\end{fmffile}
\begin{fmffile}{hh4} 	
\fmfframe(30,10)(10,10){
\begin{fmfgraph*}(100,100)
\fmfleft{i1}
\fmflabel{$\mu\nu$}{i1}
\fmfright{o1}
\fmflabel{$\rho\sigma$}{o1}
\fmf{dbl_wiggly,label=$\vec p$}{i1,v1}
\fmf{dashes_arrow,tension=1,label=$ k$,l.s=right}{v1,v1}
\fmf{dbl_wiggly,label=$\vec p$}{v1,o1}
\fmfdot{v1}
\end{fmfgraph*}
}
\end{fmffile}\newline%
\begin{fmffile}{t90} 	
\fmfframe(10,0)(10,0){
\begin{fmfgraph*}(80,80)
\fmfbottom{i1,d1,o1}
\fmflabel{$\mu\nu$}{i1}
\fmflabel{$\rho\sigma$}{o1}
\fmftop{d2,o2,d3}
\fmf{dbl_wiggly,label=$\vec p$}{i1,v1}
\fmf{dbl_wiggly,label=$\vec p$}{v1,o1}
\fmffreeze
\fmf{phantom}{o2,v2}
\fmf{dashes_arrow,label=$k$,l.s=left,left=1,tension=0.3}{v2,v3}
\fmf{dashes_arrow,l.s=right,left=1,tension=0.3}{v3,v2}
\fmf{dbl_wiggly}{v3,v1}
\end{fmfgraph*}
}
\end{fmffile}
\end{center}
\caption{One loop contributions to the graviton propagator.}%
\label{1-loop-gr}%
\end{figure}
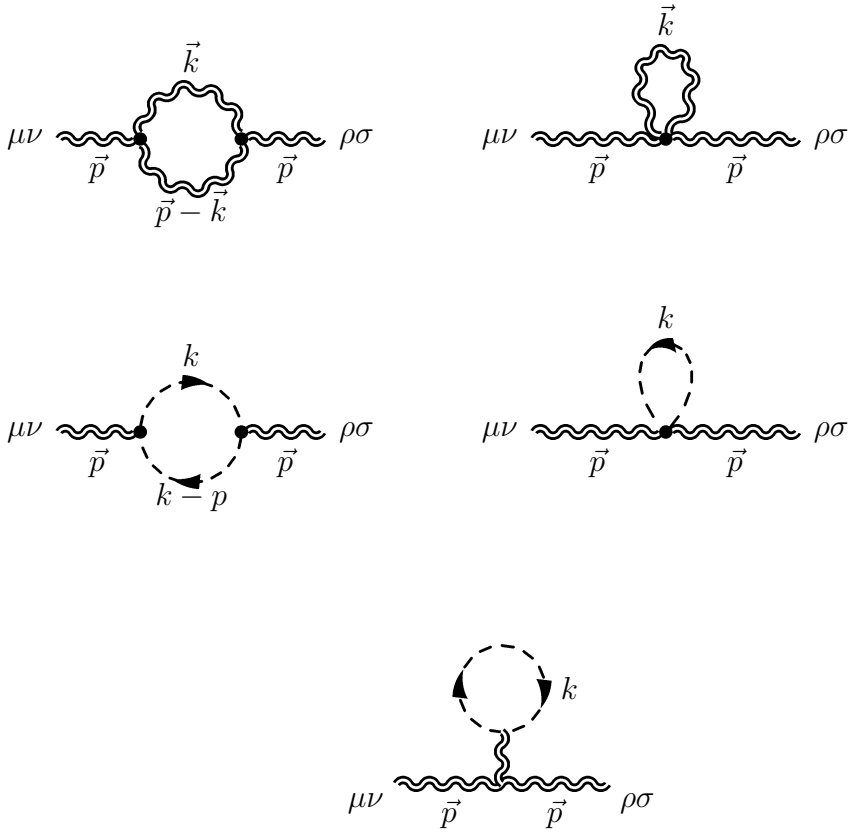

As depicted in Figure \ref{1-loop-gr}, there are five Feynman diagrams
contributing to the one loop graviton propagators. The first two are coming
from the pure Einstein gravity sector and the other three involve scalar
fields running in the loop. Since we are only interested in the $N$ dependence
of the diagrams we focus on the ones with scalar fields in the loop. In
electrodynamics, the gauge invariance enforces the photon self-energy to be
transverse. This reduces the degree of divergence from two to zero. However,
in gravity the gauge invariance does not do so. It only relates the three
diagrams that involve the scalar field in the loops. Thus the quartic
divergence, $\Lambda^{4}\delta_{\mu\nu}\delta_{\rho\sigma}$, which corresponds
to the cosmological constant term, $\Lambda^{4}\sqrt{-g}$, remains. This
difference is due to the fact that $\Lambda^{4}\sqrt{-g}$ is still
gauge-invariant whereas $m^{2}A_{\mu}A^{\mu}$ is not. This is the famous
cosmological constant problem which we are not intended to deal with here.
Next-to-leading divergent part diverges like $\Lambda^{2}$ and this is the
part that renormalizes the graviton propagator. In particular, the diagram
that involves two graviton-scalar three-vertices is
\begin{align}
I_{1}  &  =\int^{\Lambda}\frac{d^{4}k}{(2\pi)^{4}}\ \frac{i}{k^{2}-M_{a}^{2}%
}V_{3\mu\nu}^{ab}(-k,k-p;p)\frac{i}{(k-p)^{2}-M_{a}^{2}}V_{3\rho\sigma}%
^{ab}(k,p-k;-p)\nonumber\\
&  \propto N\left(  \frac{\Lambda}{M_{\mathrm{pl}}}\right)  ^{2}D_{\mu\nu
\rho\sigma},
\end{align}
as long as $M_{a}\ll\Lambda$. The diagram involving the graviton-scalar four
vertex is of the form
\begin{equation}
I_{2}=\sum_{a=1}^{N}\ \int\frac{d^{4}k}{(2\pi)^{4}}\ \frac{i}{k^{2}-M_{a}^{2}%
}V_{4}(k,-k;p,-p).
\end{equation}
The leading part of this integral is quartic in the UV cutoff and just
renormalizes the cosmological constant. The next-to-leading order is quadratic
in the UV cutoff and is proportional to $N\left(  \frac{\Lambda}%
{M_{\mathrm{pl}}}\right)  ^{2}D_{\mu\nu\rho\sigma}$, assuming that the masses
$M_{a}$ are all much smaller than the \textquotedblleft undressed (bare) UV
cutoff\textquotedblright\ $\Lambda$, which is taken to be $M_{\mathrm{pl}}$.
The last diagram in Figure \ref{1-loop-gr} has only a quartic divergence and
does not contribute to the renormalization of the graviton propagator at all.
Thus we see that the one loop graviton propagator is proportional to number of
fields $N$, as well as $\left(  \frac{\Lambda}{M_{\mathrm{pl}}}\right)  ^{2}$.

This term may be viewed as the correction to the Newton constant or
$M_{\mathrm{pl}}$. That is, the quantum gravity effects become important when
this term becomes of the same order as the classical tree level value. This
happens if the cutoff momentum $\Lambda$ is of order \footnote{Analysis of two
point function alone is not enough to deduce this result and one should also
consider graviton-scalar interaction vertex, which we will have done in the
next subsection.}
\begin{equation}
\label{reduced-cutoff}\Lambda_{\mathrm{dressed}}^{2}=\frac{M_{\mathrm{pl}}%
^{2}}{N}%
\end{equation}
which is the \emph{species dressed UV cutoff}. Besides this ``perturbative''
argument, the fact that one should use this reduced cutoff instead of
$M_{\mathrm{pl}}$ in presence of large number of species has also been backed
up by black hole physics and Hawking radiation from black holes in theories
with large number of light species \cite{Dvali:2007hz,Dvali:2008jb}.

One may also compute one loop correction to the scalar propagator. Again there
are diagrams involving only scalars and two diagrams involving gravitons. It
is immediate to see that the latter two diagrams have no parametric dependence
on the number of scalars $N$.

\subsubsection{One loop graviton-scalar vertex}

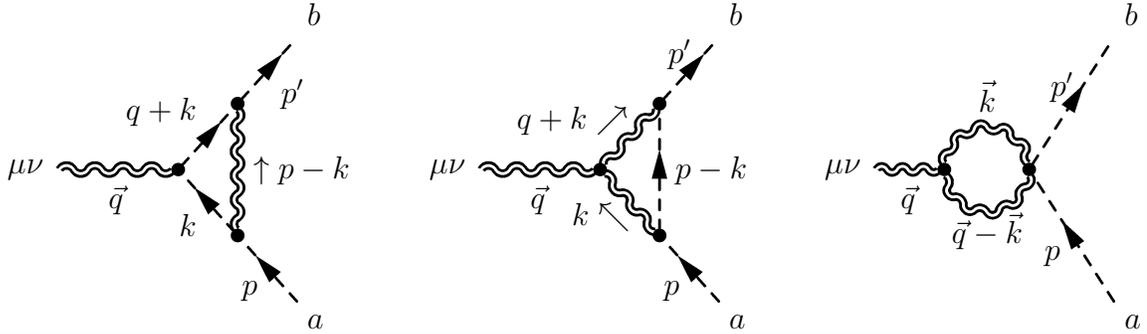
\begin{figure}[t]
\begin{center}%
\begin{tabular}
[c]{cccccccccccccccccccccccc}%
\begin{fmffile}{hpp1} \fmfframe(40,0)(40,0){ \begin{fmfgraph*}(100,100) \fmfleft{i1} \fmfright{o1,o2} \fmf{dbl_wiggly, label=$\vec q$}{i1,w2} \fmflabel{$\mu\nu$}{i1} \fmfdot{w1,w2,w3} \fmf{dashes_arrow,label=$p$}{o1,w1} \fmf{dashes_arrow,label=$k$}{w1,w2} \fmf{dashes_arrow,label=$q+k$,l.side=left}{w2,w3} \fmf{dashes_arrow,label=$p^{\prime}$}{w3,o2} \fmffreeze \fmf{dbl_wiggly,label=$\uparrow{p-k}$}{w1,w3} \fmflabel{$a$}{o1} \fmflabel{$b$}{o2} \end{fmfgraph*} } \end{fmffile} \begin{fmffile}{hpp2} \fmfframe(0,0)(0,0){ \begin{fmfgraph*}(100,100) \fmfleft{i1} \fmfright{o1,o2} \fmf{dbl_wiggly, label=$\vec q$}{i1,w2} \fmflabel{$\mu\nu$}{i1} \fmfdot{w1,w2,w3} \fmf{dashes_arrow,label=$p$}{o1,w1} \fmf{dbl_wiggly,label=$k\nwarrow$,l.d=0.01}{w1,w2} \fmf{dbl_wiggly,label=$q+k\nearrow$,l.side=left,l.d=0.01w}{w2,w3} \fmf{dashes_arrow,label=$p^{\prime}$,l.side=left,l.d=0.01}{w3,o2} \fmffreeze \fmf{dashes_arrow,label=$p-k$,l.side=right}{w1,w3} \fmflabel{$a$}{o1} \fmflabel{$b$}{o2} \end{fmfgraph*} } \end{fmffile} \begin{fmffile}{hpp3} \fmfframe(30,30)(40,0){ \begin{fmfgraph*}(100,100) \fmfleft{i1} \fmflabel{$\mu\nu$}{i1} \fmfright{o1,o2} \fmf{dbl_wiggly,label=$\vec q$}{i1,v1} \fmf{dashes_arrow,tension=0.4,label=$p$,l.s=left}{o1,v2} \fmf{dashes_arrow,tension=0.4,label=$p^{\prime}$,l.side=left,l.d=0.01}{v2,o2} \fmf{dbl_wiggly,left,tension=0.4,label=$\vec k$,l.side=left}{v1,v2} \fmf{dbl_wiggly,left,tension=0.4,label=$\vec q-\vec k$,l.side=left,l.d=0.1}{v2,v1} \fmfdot{v1,v2} \fmflabel{$a$}{o1} \fmflabel{$b$}{o2} \end{fmfgraph*} } \end{fmffile} &
&  &  &  &  &  &  &  &  &  &  &  &  &  &  &  &  &  &  &  &  &  &
\end{tabular}
\end{center}
\caption{One loop contributions to $h-\phi-\phi$ vertex. These diagrams
contribute to the corrections of graviton-scalar tree level interactions (by
renormalizing them) as well as generating the $R\phi^{2}_{a}$ term at the
level of one loop effective action. }%
\label{scalar-gravitons-exchange}%
\end{figure}

As depicted in Figure \ref{scalar-gravitons-exchange}, there are three
diagrams contributing to scalar-graviton vertex at one loop level.
Our interest in these diagrams are twofold: \emph{i)} we read the
correction to the tree level graviton-scalar three-vertex depicted in Figure 1
and, \emph{ii)} compute the coefficient of the $R\phi_{a}^{2}$ term which
appears at one loop level from these diagrams.

The details of the loop calculations, which are straightforward, are given in
the appendix, here we just quote the result. Since we are mainly interested in
the $N$ dependence of the loop expressions we only focus on that issue here.
There are no $N$ dependence appearing in any of the diagrams in Figure
\ref{scalar-gravitons-exchange}, and these diagrams, compared to the tree
level results, are proportional to $({\Lambda}/{M_{\mathrm{pl}}})^{2}$. This
in particular implies that the coefficient in front of the effective
$R\phi_{a}^{2}$ term (up to numeric factors of $1/4\pi$) is proportional to
$({\Lambda}/{M_{\mathrm{pl}}})^{2},$ and if $\Lambda$ is the species dressed
cutoff $\Lambda_{\mathrm{dressed}}$ \eqref{reduced-cutoff} this term is
suppressed by the number of light species $N$.

To summarize, the one loop correction to graviton propagator is dressed with a
power of $N$, while the graviton-scalar-scalar vertex is $N$ independent. This
result is very similar to the well-established 't Hooft $1/N$ expansion
\cite{'tHooft:1973jz}, that if we normalize the two point function to one, the
interaction term has $1/N$ suppression.

\section{$R\phi^{2}_{a}$ term in inflationary background and resolution to
$\eta$-problem}

So far we have shown that in a theory with $N$ number of species with masses
lighter than dressed cutoff $\Lambda_{\mathrm{dressed}}$
\eqref{reduced-cutoff}, the coefficient of the $R\phi_{a}^{2}$ term generated
at one loop level is $\xi/N$, where $\xi$ is an order one $c$-number. The
above analysis was carried out in a flat space background and should be
revisited for inflationary (almost de Sitter) backgrounds. It is readily seen,
however, that the basic argument behind the factors of $N$ does not depend on
the background geometry. Also, the presence of the new scale $H\ll
\Lambda_{\mathrm{dressed}}$ should not affect our argument in any
qualitatively important way. We still expect that species lighter than the
high energy cutoff $\Lambda_{\mathrm{dressed}}$ in general will contribute to
$N$. The only change concerns the very lightest species, with masses roughly
below the Hubble scale $H$, where momenta at super-Hubble scales will not
contribute, as we will now show.

We recall that the equation of motion of a free massive scalar on an inflationary
background is
\begin{equation}
\ddot{\phi}_{a}(k;t)+3H\dot{\phi_{a}}(k;t)+(\frac{k^{2}}{a(t)^{2}}+M_{a}%
^{2})\phi_{a}(k;t)=0, \label{scalar-mode-eq}%
\end{equation}
where dot denotes derivative with respect to the cosmic comoving time and
$a(t)$ is the scale factor. The relevant observation is that the modes
contributing to $N$ are the \emph{quantum modes}, e.g. those with oscillatory
(as opposed to exponentially damping or growing) behavior. To be able to solve
the above equation, let us drop the time-dependent piece $k^{2}/a(t)^{2}$ for
the moment and consider the equation
\[
\ddot{\phi}_{a}(k;t)+3H\dot{\phi_{a}}(k;t)+M_{a}^{2}\phi_{a}(k;t)=0,
\]
whose solution is of the form $\phi_{a}(k;t)=\phi_{a}^{(0)}e^{H\omega t}$ with
$\omega=-3/2\pm\sqrt{9/4-M_{a}^{2}/H^{2}}$. To have a quantum mode $\omega$
should have an imaginary part. This latter implies that $M_{a}\geq3H/2$. Note
that this result is $k$ independent and that addition of the $k^{2}/a(t)^{2}$
term will only slightly modify this result, as it is positive definite: All
the modes with $M_{a}>3H/2$, regardless of their $k$, are always quantum
modes, while modes with $M_{a}<3H/2$ are \emph{classical} for large
wavelengths (i.e. \textquotedblleft super-Hubble\textquotedblright\ physical
momenta $k/a(t)<3H/2$), and quantum mechanical for sub-Hubble momenta. Note also that the \textquotedblleft damping
coefficient\textquotedblright\ $e^{-3Ht/2}$ is removed in the process of
canonical quantization as canonical momentum conjugate to $\phi_{a}$ is
$\omega Ha(t)^{3}\phi_{a}^{(0)}$. This is
of course the standard established result in inflationary cosmic perturbation
theory and quantum field theory on curved (de Sitter) space time
\cite{Inflation-Books}.

Since we are only interested in the UV behavior of the loop integrals,
we can instead of integrating over $k$ all the way from zero to $\Lambda
_{\mathrm{dressed}},$ restrict the integral to go from $H$ to $\Lambda
_{\mathrm{dressed}}$. In this way we avoid the unnecessary complication with
super-Hubble modes. In summary, all the modes lighter than $\Lambda
_{\mathrm{dressed}}$, with both super-Hubble or sub-Hubble masses, contribute
to the $N$ in the loop integral. In other words, as long as $H\lesssim
\Lambda_{\mathrm{dressed}}$, $N$ is the same for inflationary and flat space and
\[
\Lambda_{\mathrm{dressed}}^{2}=\frac{M_{\mathrm{pl}}^{2}}{N},
\]
where $N$ is the number of species lighter than the cutoff, $\Lambda
_{\mathrm{dressed}}$. In particular, the coefficient of the $R\phi_{a}^{2}$
term generated at one loop will become $\xi/N$, with $\xi$ of order one.

We are now ready to address the $\eta$-problem. To this end, we recall that
the one loop corrected action is
\begin{equation}
L=L_{cl}+\frac{\xi}{N}R\phi_{a}^{2}\,.
\end{equation}
Hence, the slow-roll parameter $\eta_{ab}\equiv M_{\mathrm{pl}}^{2}%
\frac{V_{ab}}{V}$, where $V_{ab}=\frac{\partial^{2}V}{\partial\phi_{a}%
\partial\phi_{b}}$, is%
\begin{equation}
\eta_{ab}=\eta_{ab}^{cl}+\frac{\xi}{N}\frac{R}{3H^{2}}\delta_{ab}\simeq
\eta_{ab}^{cl}+\frac{4\xi}{N}\delta_{ab}\,.
\end{equation}
To have a successful slow-roll inflationary period we usually demand $\eta
\sim0.01$, and if we assume $\eta^{cl}\sim0.01$, quantum corrections to $\eta$
will be suppressed enough for $N\gtrsim100$.\footnote{To complete this
discussion we note that in multi-filed inflationary models, like N-flation
\cite{Nflation} or M-flation \cite{Mflation}, there is a certain combination
of the fields which plays the role of effective inflaton and the original
\textquotedblleft physical\textquotedblright\ field should be rescaled with
appropriate powers of number of fields $N$ so that this effective inflaton
field is canonically normalized. The $R\phi_{a}^{2}$ term, being quadratic in
the $\phi_{a}$'s, will not receive any normalization factors due to the $N$
scaling relating canonically normalized inflaton and the original fields.}

\section{Discussion}

The bottom-up $\eta$-problem seems quite generic to all models of inflation that
involve a scalar field minimally coupled to gravity. Even if the parameters
of the inflaton potential are chosen meticulously at tree-level, the loop
corrections that arise from interactions of the graviton with the scalar field
create the quadratically divergent conformal mass term which leads to the
$\eta$-problem, if the UV cutoff of the theory is of order $M_{\mathrm{pl}}$.
This kind of $\eta$-problem is of course different from the ``top-down'' $\eta$-problem  arising
within the string theory setups in which the volume modulus stabilization
often resurrects the $\eta$-problem. The precision that should be enforced upon the tree-level parameters are often not needed to sustain inflation, but to match the observed density of perturbation \cite{arXiv:1001.4538}. In this letter we examined the
possibility of circumventing the $\eta$-problem, in the former sense discussed
above, in many-field models of inflation that are minimally coupled to
gravity. These many-fields whose masses are assumed to be smaller than $\mpl/\sqrt{N}$, have a natural appearance in effective low energy field theory description of quantum gravity models. As we argued $N\gtrsim 100$ will resolve the $\eta$-problem. Even though it is not necessary for our argument, the scalar fields should be non-interacting to realize inflation.

 One example of such many-field models is N-flation \cite{Nflation} which has the $O(N)$ symmetric potential $V=\sum_{i=1}^{N}\frac{1}{2}m^2\phi_i^2$. As stated above, a few hundred scalar fields will be enough to circumvent the $\eta$-problem. Like its chaotic counterpart, the mass parameter $m$ has to be $\sim 10^{-6}\mpl$ which is smaller than $\Ld$ unless one resorts to an unnaturally large number of scalar fields, \textit{i.e.} $N\sim 10^{12}$. This of course comes at the price of  exposing the model to quantum instability of the type discussed in \cite{gauged-M-flation}. Namely, the quantum fluctuations of these light fields may dominate over the classical evolution of the inflaton. Assisted model with quartic potential $V=\sum_{i=1}^{N}\frac{1}{2}\lambda\phi_i^4$ is another possibility. To have an observationally viable  model, the effective coupling must be around $10^{-14}$ and thus with $\lambda\simeq 1$, one needs around $10^{14}$ scalar fields. This scenario also suffers from the above quantum instability with such a large number of massless scalar fields. Another disadvantage of both these two models is that the physical excursion of the fields is larger than $\Ld$ during the required $60$ e-folds of inflation.

Another explicit example is M-flation \cite{Mflation} or its gauged  version \cite{gauged-M-flation} where the inflaton potential is constructed by three $N\times N$ non-commutative hermitian matrices whose action is invariant under $U(N)$. The
classical dynamics is simplified considerably in the $SU(2)$ sector where
these three scalar fields are proportional to the generators of the $SU(2)$
algebra. Gauged M-flation, in addition to the above ingredients, has an extra
$U(1)$ field, associated with ``center of mass '' $U(1)\in U(N)$. M-flation in the $SU(2)$ sector besides the inflaton field contains some number of ``spectator fields'' which do not contribute to classical inflationary trajectory while can be excited quantum mechanically and appear in the loops. For the gauged M-flation there are $2N^2+1$ such scalar modes and $3N^2-1$ massive vector modes. These modes have a hierarchical spectrum, {\it i.e.} they can be lighter or heavier than the Hubble scale $H$, for the explicit masses see \cite{gauged-M-flation}. Not all these modes are light enough to be counted in the dressed cutoff. As discussed in \cite{gauged-M-flation}, the number of ``contributing species'' $N_s$ varies between $3\times10^{5}$
and $10^{6}$, depending on the region of the potential inflation happens. Hence, the species dressed UV cutoff is $10^{-3}M_{\mathrm{pl}%
}\lesssim\Ld\lesssim5\times10^{-3}M_{\mathrm{pl}}$. Consequently the
$R\phi^{2}$ term is suppressed by a factor of $\lesssim10^{-5}$ and could be
safely ignored. Gauged M-flation could be motivated from the branes dynamics in an appropriate flux where
the above matrices correspond to three of the perpendicular directions of a stack of $N$
D3-branes which are scalars in the adjoint representation of the $U(N)$. As such, although the ``quantum" $\eta$-problem is resolved for M-flation, embedded within string theory, one should still deal with the ``stringy $\eta$-problem'' (cf. introduction for further references and discussions).

Finally, the term leading to the $\eta$-problem, $R\phi_a^2$, is a one-loop but marginal operator. One may naturally worry about other loop corrections, that enhancement factors of $N$ will dominate over ${\Ld}/{\mpl}$ suppressions. It is straightforward to show that the largest such $N$ enhancement factor appears in the graviton two point function (at higher loops) for which this factor is $\left(N{\Ld^2}/{\mpl^2}\right)^l$, where $l$ is number of loops. All the other diagrams will be of the form $N^k(\frac{\Ld^2}{\mpl^2})^{l}$ with $k<l$. Hence our one loop result seems to be also valid to all orders.

\section*{Acknowledgements}

A.A. was supported by the G\"{o}ran Gustafsson Foundation. U.D. is supported
by the Swedish Research Council (VR) and the G\"{o}ran Gustafsson Foundation.
We would like to thank Liam McAllister for helpful discussions.

\appendix

\section{Details of the loop calculations}

Now let us focus on all the diagrams which may generate the $R\phi^{2}$ term,
or equivalently $\frac{1}{M_{\mathrm{pl}}} (\partial^{2} \hat{h})\phi^{2}$, at
the one-loop order. The first one, is given by the left diagram in Figure
\ref{hphiphi1}, where a graviton exchange between two external scalars
modifies the scalar form factor. The diagram is proportional to
\begin{align}
\phi(p)\delta h^{\mu\nu}(q)\phi(p^{\prime})  &  =\int\frac{d^{4} k}{(2\pi
)^{4}}\left(  \frac{1}{M_{\mathrm{pl}}} \left[  p_{\alpha}k_{\beta}-(p\cdot k)
\eta_{\alpha\beta}\right]  \frac{i(\eta^{\alpha\gamma}\eta^{\beta\kappa}%
+\eta^{\alpha\kappa}\eta^{\beta\gamma}-\eta^{\alpha\beta}\eta^{\gamma\kappa}%
)}{(p-k)^{2}+i\epsilon} \right. \nonumber\\
&  \left.  \frac{i}{k^{2}+i\epsilon} \frac{1}{M_{\mathrm{pl}}} \left[
p^{\prime}_{\gamma}(k+q)_{\kappa}-\eta_{\gamma\kappa}p^{\prime}\cdot
(k+q)\right]  \frac{1}{(k+q)^{2}+i\epsilon}\right. \nonumber\\
&  \left.  \frac{1}{M_{\mathrm{pl}}} \left[  k_{\mu}(k+q)_{\nu}-\eta_{\mu\nu
}k\cdot(k+q)\right]  \right)  .
\end{align}
What we are interested in is the divergent part of the above integral which
multiplies the generated $R\phi^{2}$ term. There are six momenta in the
numerator, only four of which are internal. Thus the integral will be
divergent as
\begin{equation}
\label{leading-contribution}\frac{1}{M_{\mathrm{pl}}}{\left(  \frac{\Lambda
}{M_{\mathrm{pl}}}\right)  }^{2}.
\end{equation}

The other diagram that contributes to the $h-\phi-\phi$ vertex is the middle
one in Figure \ref{scalar-gravitons-exchange}. To estimate the leading
divergent part of this diagram, one should note that the vertex that involves
three $\hat{h}$ is proportional to $1/M_{\mathrm{pl}}$. More specifically,
\begin{align}
W_{3\alpha\beta\mu\nu\gamma\kappa}  &  =\frac{1}{8M_{\mathrm{pl}}}%
\sum_{\mathrm{sym}} \left[  -4 \eta_{\mu\gamma}\eta_{\kappa\alpha}(p_{2}%
.p_{3})+2\eta_{\gamma\kappa}\eta_{\mu\alpha}\eta_{\nu\beta} (p_{2}.p_{3}%
)-\eta_{\gamma\kappa}\eta_{\mu\nu}p_{2{\alpha}}p_{3{\beta}}+2\eta_{\mu\gamma
}\eta_{\eta\nu} p_{2\alpha}p_{3\beta}\right. \\
&  +\left.  4\eta_{\gamma\alpha}\eta_{\nu\beta}p_{2\mu}p_{3\eta}\right]  \,
.\nonumber
\end{align}
Sum is over symmetrization on the index pairs $(\alpha,\beta)$, $(\gamma,
\kappa)$ and $(\mu,\nu)$ and also the six permutation done over momentum index
triplets $\alpha\beta p_{1}$, $\gamma\kappa p_{2} $ and $\mu\nu p_{3}$. Again,
this diagram has six momenta in the numerator, two of which are external. Thus
the divergent part of diagram behaves as \eqref{leading-contribution} times
the terms among which $R\phi^{2}$ exists.

Finally, let us look at the right one-loop diagram in Fig.
(\ref{scalar-gravitons-exchange}). The vertex $\hat{h}\hat{h}\phi\phi$ is
proportional to $1/M_{\mathrm{pl}}^{2}$. In more details:
\begin{align}
V^{ab}_{4\alpha\beta\mu\nu}  &  =\frac{1}{M_{\mathrm{pl}}^{2}}\left[
\eta_{\mu\nu} p_{1\alpha}p_{2\beta}-2\eta_{\beta\nu}p_{1\alpha}p_{2\mu}%
+\eta_{\mu\nu} p_{1\beta}p_{2\alpha}-2\eta_{\alpha\nu} p_{1\beta}p_{2\mu}%
+\eta_{\mu\nu}p_{1\beta}p_{2\alpha}\right. \\
&  - \left.  2\eta_{\alpha\nu} p_{1\beta} p_{2\mu} - 2\eta_{\beta\mu}
p_{1\alpha}p_{2\nu}+\eta_{\nu\mu}p_{1\beta}p_{2\alpha}-2\eta_{\alpha\mu
}p_{1\beta}p_{2\nu}\right]  \,.\nonumber
\end{align}
The diagram contains two propagators in the denominator and four momenta in
the numerator, two of which are external. Thus the leading correction will be,
again, of order \eqref{leading-contribution} times the terms among which
$R\phi^{2}$ exists. Note that besides the conformal mass term, there are also
other higher dimensional operators, whose explicit coefficients could be
obtained by exactly calculating the amplitudes. For example, from usual
tensorial analysis, terms like $R^{\mu\nu}\partial_{\mu}\phi\partial_{\nu}%
\phi$ are expected to be generated from such one loop diagrams. However, all
these terms are suppressed by extra powers of the cutoff and also by slow-roll
parameters in an inflationary background.

The other point which is worth mentioning is that the existence of higher
order self-interactions for the scalar fields will not disturb our argument.
For example inclusion of mass term in the potential for the inflaton, will
modify the scalar field propagator and also add corrections proportional to
$M^{2}_{a}/M_{\mathrm{pl}}$ in the $\hat{h}-\phi-\phi$ vertex. Such terms
would at most introduce correction of order $\frac{M_{a}}{\Ld}{\left(
\frac{M_{a}}{M_{\mathrm{pl}}}\right)  }^{2}\frac{1}{M_{\mathrm{pl}}}$ to the
leading contribution \eqref{leading-contribution}. Thus, as long as
$M_{a}\lesssim\Ld\ll M_{\mathrm{pl}}$, the effect of such terms are very
small. Other forms of potential for the scalar field induce vertices that lead
to more loops whose effect is more suppressed in comparison with the one-loop diagrams.

\end{document}